# Linguistic-Pattern-Based Optimization of Economic and Spatial Uniformity Criteria in Facility Layout Problems

Jerzy Grobelny[1] [0000-0001-9791-9395], Rafał Michalski[1*] [0000-0002-0807-1925]

[1]Faculty of Management,
Wrocław University of Science and Technology
27 Wybrzeże Wyspiańskiego, 50-370 Wrocław, Poland
jerzy.grobelny@pwr.edu.pl,
rafal.michalski@pwr.edu.pl

*Corresponding author

**Abstract:** This paper extends prior work on linguistic-pattern-based facility layout optimization by enhancing the LP-Alinks framework with an explicit spatial-uniformity criterion. While earlier studies demonstrated that linguistic patterns can effectively encode expert knowledge and guide agent-based layout emergence, their optimization scope remained limited to cost-oriented objectives. To address this gap, we introduce the Normalized Coverage Score (NCS), a scale-adjusted measure of spatial evenness that complements the classical flow–distance economic objective and enables systematic exploration of cost–uniformity trade-offs. A full factorial experiment comprising 324 conditions evaluates the combined effects of problem size, link density, virtual-force scaling, and two families of membership functions on both objectives. Five-way and nested three-way ANOVAs reveal that structural factors (number of objects and link density) exert a dominant baseline influence on both economic performance and spatial uniformity. However, while economic cost is overwhelmingly governed by these structural characteristics, uniformity is substantially more sensitive to the linguistic-pattern parameters, which control the dispersion–compaction dynamics of the emerging layouts within the structural constraints. Based on these interactions, we derive a parameter-selection matrix that prescribes settings for cost minimization, uniformity maximization, or balanced compromise. Comparative analyses with Drezner's method, MDS, and non-metric MDS demonstrate that the extended LP-Alinks framework consistently attains superior uniformity while maintaining competitive economic performance, making it a robust and interpretable decision-support tool for early-stage facility layout design.

**Acknowledgement**: The work was partially financially supported by the Polish National Science Centre grant no. 2019/35/B/HS4/02892: *Intelligent approaches for facility layout problems in management, production, and logistics*.

## 1. Introduction

Designing effective facility layouts requires balancing ergonomic principles with computational constraints. Classical ergonomics and human–system interaction research (McCormick, 1976; Wierwille, 1984), emphasize that high quality layouts must satisfy both first order criteria, such as the importance and frequency of use of individual components, and second order criteria describing their functional and operational relationships, such as order of use or task dependent adjacency. These requirements reflect the broader ambition of designing workspaces that are safe, efficient, intuitive, and resistant to human error.

In practice, these design challenges are typically formalized as Facility Layout Problems (FLP), often expressed through flow–distance formulations. However, classical FLP models are computationally demanding and rely on restrictive assumptions about discrete placement, which limits their usefulness in early-stage design.

To support more flexible spatial exploration, scatter-plot-based methods were introduced as continuous alternatives that visualize relational structure without assigning objects to predefined locations. Deterministic formulations, e.g., Drezner (1980, 1987a), and fuzzy and stochastic-based variants (Grobelny, 1987a, 1987b, 1999) can reveal meaningful adjacency patterns, but their geometric analogies offer limited means for incorporating expert reasoning beyond link intensities. Recent developments address this limitation by introducing linguistic patterns (LPs) as a mechanism for expressing qualitative spatial requirements in a form close to natural language (Grobelny & Michalski, 2022, 2025).

By incorporating stochasic features, the LP-Alinks approach generates multiple diverse scatter-plot layouts, with their form depending on the algorithm parameters. These layouts are not intended as final designs but rather as decision-support tools that help reveal promising spatial relationships, highlight potential conflicts, and provide informed starting points for subsequent refinement. Moreover, LPs formulations naturally support multicriteria optimization, accommodating both the traditional economic objective function (EGF) based on link–distance products, and additional spatial criteria that manifest visually in the resulting scatter plots.

In such context, the main objective of this study is to develop and evaluate an extended LP-Alinks framework that simultaneously optimizes economic efficiency and spatial uniformity in FLPs. In particular, we aim to formalize spatial uniformity as an explicit optimization criterion and to investigate how algorithmic and structural factors influence the trade-off between cost and uniform layout structure. The main contributions of this research are:

- We augment the classical flow–distance formulation by incorporating an explicit spatial uniformity objective, enabling the simultaneous pursuit of cost-efficient and geometrically interpretable layouts.
- We introduce a grid-based uniformity index, called Normalized Coverage Score (NCS), that quantifies the dispersion of objects across a discretized plane. The measure reflects the proportion of occupied grid cells, provides a transparent operationalization of layout readability, and mitigates the clustering effects common in classical layout generation techniques.
- Using a full factorial experimental design with 324 conditions, we investigate how key problem characteristics, that is, number of objects, link density, distance- and link-membership functions, and virtual-force scaling parameter, shape the behavior of LP-Alinks with respect to both economic cost and spatial uniformity.

- We benchmark LP-Alinks against Drezner's deterministic scatter-plot method, classical multidimesional scaling MDS (Torgerson, 1952, 1958) and, and non-metric MDS, NmMDS (Kruskal, 1964a, 1964b; Shepard, 1962a, 1962b). This analysis offers a comprehensive examination of cost–uniformity trade-offs and demonstrates conditions under which LP-Alinks outperforms or complements these traditional techniques.
- Based on factorial simulation results, we derive a parameter-selection matrix linking problem settings to configurations that best optimize economic performance, maximize spatial uniformity, or provide a balanced compromise between the two objectives.

Together, these contributions position LP-Alinks as a human centered operational research framework that integrates expert reasoning with computational optimization, illuminates the trade offs between economic performance and spatial clarity, and generates interpretable design alternatives suited for early stage facility planning.

The remainder of this paper is organized as follows. Section 2 presents the theoretical background and related research. It first discusses the FLP and then reviews the issue of spatial uniformity, including related studies and the proposal of the Normalized Coverage Score. Section 3 describes the simulation experiment, including the experimental design, procedure and the results of the study. Section 4 discusses the findings, interprets the recommendations, compares the proposed approach with other algorithms, and outlines linitations and future research. Finally, Section 5 concludes the paper and summarizes its contributions.

# 2. Theoretical background and related research

## 2.1. Facility Layout Problem (FLP)

The FLP classically concerns arranging a set of objects in a given space so as to minimize total flow–distance cost, typically expressed as:

$$\min_{i<j} \sum_{i<j} \lambda_{o_i o_j} \Delta\left(o_i, o_j\right)$$

where $\lambda_{o_i o_j}$ represents the strength of proximity between objects $o_i$ and , $o_j$, and $\Delta\left(o_i, o_j\right)$ captures the distance between them. This formulation underpins both industrial and human–computer interaction contexts (Drira et al., 2007; Tompkins et al., 2010).

A major stream of FLP research assumes a finite, predefined set of possible discrete locations. When all objects are equal in size and must be assigned to distinct locations, the problem becomes the Quadratic Assignment Problem (QAP). Originating in the seminal work of Koopmans & Beckmann (1957), the QAP models assignment costs in terms of pairwise flows and distances between sites. While powerful conceptually, QAP and related formulations are known to be strongly NP-hard (Sahni & Gonzalez, 1976), making exact optimization infeasible for all but the smallest instances. Consequently, a rich family of approaches has emerged, including various types of heuristics, metaheuristics, biologically inspired algorithms, and hybrid approaches (Burkard, 2013; Pitsoulis & Pardalos, 2008). Recent developments even explore machine-learning-based solvers, such as deep reinforcement learning approaches tailored to the Koopmans–Beckmann formulation (Li et al., 2026; Unger & Börner, 2021).

Despite this diversity, most classical FLP methods implicitly constrain spatial structure by restricting layouts to predefined grids or discrete location sets. Such assumptions are too limiting for early stage design, when practitioners often need to explore non regular, emergent arrangements before applying stricter feasibility constraints. To overcome these limitations, another trend of studies allows arbitrary facility dimensions. In these unequal-area facility layout problems, the assumption of fixed, identical sites is removed, providing greater flexibility for realistic engineering projects (Meller & Gau, 1996). Research in this domain includes mixed-integer programming, slicing-tree representations, and advanced disjunctive constraints for preventing overlap. Enhanced models, such as those of Sherali et al. (2003) or Jankovits et al. (2011), provide more accurate approximations of area and adjacency constraints. Extensions of these ideas support multi-objective models involving cost, closeness ratings, and makespan (Bhuiyan et al., 2021).

In many practical FLP applications, designers use scatter-plot visualizations to express relational patterns based on closeness matrices. Algorithms that aim at optimizing solutions without strictly assigning objects to predefined positions were proposed by Drezner (1980, 1987b) or Anjos & Vannelli (2002). Within this trend, qualitatively different proposals involving fuzzy and stochastic components were introduced by (Grobelny, 1987b, 1987a, 1999). Here, the objects move based on virtual forces between them. Attractive forces draw related components together, while dispersive forces prevent excessive crowding, resulting in a variety of plausible layouts that preserve relational structure. These visual layouts may also not represent final, feasible floor plans, but instead serve as initial solutions or design cues for subsequent optimization. Empirical studies show that scatter-plot–generated initial solutions improve the effectiveness of classical algorithms such as CRAFT and simulated annealing (Grobelny & Michalski, 2017, 2020).

Recent developments expand this idea by introducing linguistic patterns (LPs) into the object (agent) based simulation of layout emergence (Grobelny & Michalski, 2022, 2025). LPs encode desired spatial relations such as "*strongly related objects should be close*" or "*objects should not overlap*", as logical expressions in a form reminiscent of natural language. Their fulfillment is measured using Łukasiewicz implication and equivalence relations, providing a continuous notion of "truth deficit." These deficits generate attraction or repulsion forces governing the motion of objects during simulation. This enables designers to articulate requirements using intuitive linguistic rules, thereby incorporating expert knowledge, imprecise preferences, and qualitative reasoning into the optimization process. LP-based scatter plots provide diverse, interpretable proposals and naturally support multi-criteria optimization, including cost-based and spatial criteria.

In manufacturing and logistics, FLP plays a central role in reducing material-handling costs, which can constitute a large share of operating expenses. Effective layouts can deliver significant savings and increase throughput (Drira et al., 2007; Tompkins et al., 2010). Beyond industrial contexts, FLP models have been adapted to optimize human–machine interfaces and control panel design, situations where minimizing physical or visual travel supports improved usability. Studies show that regular-grid FLP algorithms can meaningfully optimize interface layouts based on frequency, sequence of use, and criticality (Grobelny et al., 1995; Grobelny & Michalski, 2015). These approaches complement established usability heuristics, such as Nielsen's principles, by providing quantitative models of spatial efficiency.

The closeness values $\lambda_{o_i o_j}$ used in FLP models can reflect diverse criteria: material flows, adjacency desirability, order of use, ergonomic constraints, safety requirements, or eye-movement metrics in HCI contexts. Moreover, distance measures may shift from Euclidean or Manhattan metrics in shop floors to behavioral metrics related to interface design. Modern FLP research often extends the classical flow-distance objective to address sustainability, flexibility, robustness, and operator well-being. This includes multi-criteria decision-making frameworks in both logistics and interactive-system design (Bonney & Williams, 1977; McKendall & Hakobyan, 2021; Tompkins et al., 2010).

FLP research therefore spans discrete and continuous optimization, deterministic and stochastic models, and increasingly, semantic and knowledge-driven mechanisms. The LP-Alinks framework builds on these developments by integrating linguistic constraints with virtual-force simulation, allowing systematic exploration of cost–uniformity trade-offs.

## 2.2. Uniformity of objects spatial distribution

The classical objective in the FLP is the minimization of the total material-handling cost, typically expressed as the sum of link intensities multiplied by inter-departmental distances. However, recent research in spatial optimization and visualization underscores that geometric quality and spatial regularity are equally important for interpretability, usability, and design processes. The literature consistently demonstrates that relying solely on the economic objective can lead to highly irregular or congested layouts that are difficult to interpret and suboptimal from ergonomic, operational, or spatial-planning perspectives.

### 2.2.1. Related research

The classical distance–flow objective promotes economic efficiency but often leads to spatially collapsed configurations, as confirmed in the literature on eigenvector-based scatter-plot layouts (Grobelny, 1999). Grobelny & Michalski (2020) although focused on initialization effects, cites and reiterates this findings confirming that Drezner-style eigenvector layouts can be considerably misleading. Similar qualitative observations reinforced by empirical evaluations were confirmed in their further studies (Grobelny & Michalski, 2022, 2025),

Work in visualization science shows that general scatter-plot-based representations often suffer from uneven sample distribution, which reduces readability, increases local clutter, and impairs interpretation. Rave et al. (2024) document that point-based layouts tend to form dense clusters that "*occlude structure*," and propose sector-based transformations to deliberately increase sample uniformity, evaluating the improvements using explicit uniformity measures. This work confirms that uniform spatial distribution is a measurable and desirable geometric property, and that improving it materially affects readability.

A closely related perspective comes from spatial optimization. Zhou & Murray (2025) introduce the concept of spatial configuration efficiency, showing that real facility systems often deviate from idealized, regularly distributed lattices, and propose a formal optimization model to quantify how existing facilities diverge from a uniform reference pattern. Their argument is that regularity itself is a criterion of spatial efficiency, especially when layouts must be comprehensible, extendable, or scalable.

Within FLP-specific research, non-classical objectives have been proposed to overcome the deficiencies of purely economic optimization, though many studies underscore the importance of treating FLP multidimensionally. For example, Song et al. (2008, 2023) incorporated layout balance, accessibility, and functional constraints into a multi-objective optimization framework, acknowledging that operational usability depends on more than cost minimization. García-Hernández et al. (2015) introduced an interactive multi-objective genetic algorithm that allows experts to inject subjective preferences about geometric and qualitative layout properties, further reinforcing that layout quality is inherently multidimensional.

Contemporary reviews (Pérez-Gosende et al., 2021) highlight that while FLP research has explored many alternative criteria, such as adjacency satisfaction, shape compliance, safety, accessibility, ergonomics, and robustness, measures of spatial evenness or dispersion are still missing from standard multi-objective formulations. In multi-objective FLP frameworks, the need for complementary criteria is already well established, but these works focus predominantly on functional or ergonomic constraints rather than geometric regularity.

Outside FLP, recent work demonstrates that spatial uniformity can be rigorously formalized and quantified using geometric or statistical constructions on the (hyper)sphere. Some of the proposals in this regard involve analyses based on inter-point distance distributions on the sphere (Del Bono et al., 2024) or projection-based uniformity tests (García-Portugués et al., 2023) to diagnose how empirical point sets deviate from an ideal uniform process. This line of research reinforces that uniformity is not merely a visual or heuristic property but a well-defined geometric–statistical construct supported by analytical tools.

The publications identified above demonstrate the importance of quantifying uniformity but do not provide a ready-to-use, FLP-specific metric. Visualization research Rave et al. (2024) measures uniformity in scatter-plots, but their metrics are tied to pixel density and visual de-cluttering, not physical layout optimization. Spatial equity work (Chen, 2023) proposes measures of equitable access, which capture distributional fairness but not geometric dispersion per se. Spatial-configuration efficiency (Zhou & Murray, 2025) compares facilities to an ideal lattice, but their method focuses on global coverage patterns, is computationally heavy, and is not embedded in FLP simulation workflows. Mathematical uniformity measures (Del Bono et al., 2024; García-Portugués et al., 2023) provide powerful theoretical tools, but they operate in normalized high-dimensional spaces and are not directly interpretable for grid-based manufacturing layouts.

Together, these considerations shows that uniformity is a scientifically recognized dimension of quality in spatial configurations, yet it remains largely absent in mainstream FLP optimization. Despite advances, none of the surveyed methods offer a simple, domain-tailored metric for quantifying the evenness or dispersion of object placement on an unstructured 2D design plane. Existing approaches either evaluate uniformity indirectly (via functional performance), or operate in contexts not directly compatible with FLP simulation (e.g., pixel-based scatter-plot measures or theoretical distributions).

The absence of a practical, interpretable, and computationally efficient uniformity measure for FLP motivated us to propose a new metric.

#### 2.2.2. New Spatial Uniformity Measure Proposal

To evaluate geometric quality of the generated layouts in terms of uniformity, we introduce a Normalized Coverage Score (NCS), which quantifies the proportion of the discretized design plane that is meaningfully utilized by the layout.

$$\text{Normalized Coverage Score} = \frac{\mathit{Number\ of\ occupied\ cells\ of\ the\ regular\ grid}}{\left\lceil\sqrt{\mathit{Number\ of\ objects\ in\ a\ problem}}\right\rceil^2}.$$

Formally, the index can be described in the following way. Let the layout be represented by the set of object locations. Let $N$ denote the number of objects in the layout and

$$\mathcal{O} = \{o_1, \dots, o_N\} \subset R^{2}$$

their coordinates in the design plane. Let us superimpose the layout of these objects on the design plane, which is a regular square grid ($\mathcal{G}$)

$$\mathcal{G} = \lceil\sqrt{N}\rceil \times \lceil\sqrt{N}\rceil,$$

with

$$K = \lceil\sqrt{N}\rceil^2$$

being the number of cells of equal area. Let

$$\mathcal{G} = \{g_1, \dots, g_K\}$$

be the cells. For $g_k$ cell, define the occupation indicator in the following way:

$$I(g_k) = \begin{cases} 1, & \text{if at least one } o_i \in \mathcal{O} \text{ lies in } g_k, \\ 0, & \text{otherwise}. \end{cases}$$

Then, the define a NCS as the fraction of occupied cells in grid $\mathcal{G}$:

$$\text{NCS} = \frac{1}{K} \sum_{k=1}^{K} I(g_k) \in [0, 1]$$

Higher values of NCS indicate more evenly dispersed layouts. NCS = 1 corresponds to each cell containing at least one object, while low values indicate clustering. This formulation allows to operationalize how effectively the available space is utilized by the layout.

Our proposal exhibits three theoretical properties involving normalization, coverage, and spatial efficiency. The denominator $K$ grows with $\lceil \sqrt{N} \rceil^2$, providing scale invariance, that is, when problem sizes increase, the grid grows smoothly, maintaining comparability of the NCS score. This aligns well with the Farmer et al. (2019) postulate that uniformity measures should be invariant to sample size. The score directly measures how much of the design area is effectively used, which is aligned with the concept of regularized scatter-plot coverage in visualization research (Rave et al., 2024). Additionally, the metric formalizes the notion suggested by Zhou & Murray (2025) that high-quality spatial configurations approximate regular spatial coverage patterns, i.e., near-lattice distributions.

Thus, the proposal is theoretically aligned with contemporary research on uniform spatial distributions, computationally simple, and intuitive for engineers and ergonomists, as it directly measures 2D space coverage, unlike classical cost-based or adjacency metrics. It is also invariant under permutations of object labels and compatible with both continuous scatter-plot layouts and grid-based FLP frameworks.

Given the evidence that uniformity improves readability, reduces clutter, and supports human-centered decision-making, this measure provides a practical operationalization of those principles within FLP. Interpreting uniformity as a NCS positions the metric as a geometrically principled, size-adjusted indicator of spatial quality, complementing the classical economic objective.

## 3. Simulation Experiment

The simulation experiment was designed to systematically evaluate the behavior and performance of the LP-Alinks approach in generating scatter-plot-based facility layouts that simultaneously satisfy two distinct optimization criteria. The first is the classic economic objective function, defined as the total sum of link–distance products across all object pairs, prioritizing the minimization of flow costs and handling effort. The second is the NCS introduced in the previous section, serving as an indicator of uniform object distribution, congestion avoidance, load balancing, and overall spatial sustainability. While cost minimization remains central, the importance of well-distributed layouts has increased in modern manufacturing, where ergonomics and system clarity are essential for sustainable operation.

Because the LP-Alinks method's virtual-force mechanism, grounded in linguistic patterns, is inherently sensitive to both problem structure and algorithm parameterization, a full factorial design was employed. This enabled a comprehensive assessment of individual effects and the nonlinear interactions between variables. Previous studies (Grobelny & Michalski, 2025) and further preliminary investigations indicated that factors such as link density, membership-function curvature, and force scaling can have synergistic or antagonistic influences on performance. Consequently, the study examines how these parameters, along with the number of objects and link density, affect the system's ability to achieve high-quality solutions across diverse, randomized scenarios, ensuring that results generalize beyond specific link-matrix structures.

Beyond measuring algorithmic effectiveness, the experiment sought to derive empirically grounded guidelines for parameter selection and provide practical recommendations for real-world applications.

## 3.1. Experimental Design and Procedure

### 3.1.1. Design Overview

To rigorously assess the effects and interactions of key algorithmic and problem-specific parameters, the experiment employed a full factorial within-subjects design. Five independent factors were manipulated:

- Number of objects (NO): 15, 25, or 35 items.
- Random relative link density (LD): 10%, 40%, or 70% of the maximum link density, calculated as the number of links in a fully connected network where every object is linked to every other.
- Single step object displacement magnitude coefficient for maximum virtual force (VF): 1%, 3%, or 5% of the plane's maximum dimension.
- Small distance-related membership function (MD): four variants determining the rate at which membership decreases with distance. The applied functions are illustrated in Fig. 1.

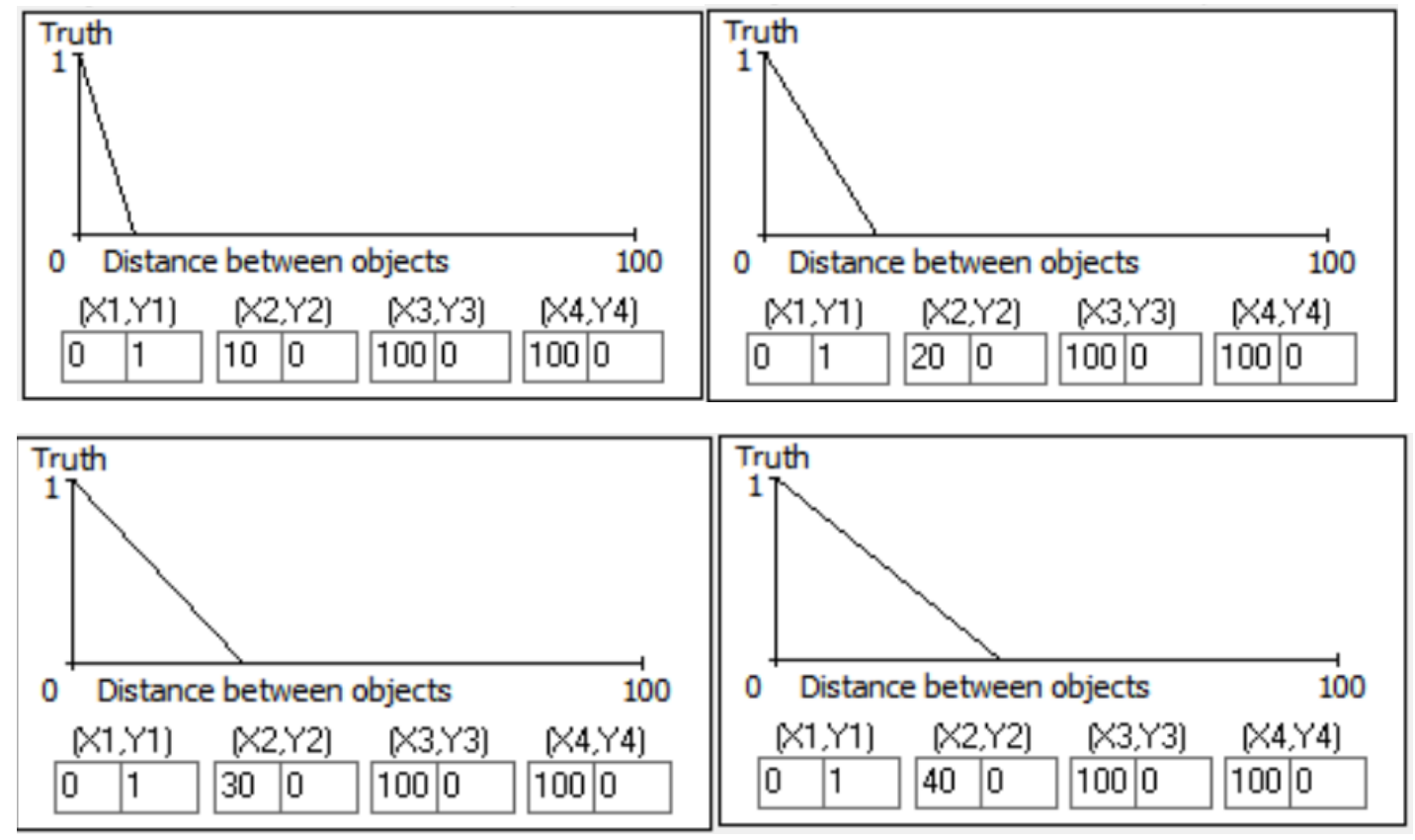

**Fig. 1.** Definition of the membership functions (from the left MD10, MD20, MD30, MD40) for the fuzzy variable *Small_Distance* in the universe of discourse *Percent of the maximum Manhatan distance between objects*.

- Link-value membership function (ML) type: *Concave*, *Linear*, or *Convex*. These function shapes are presented in Fig. 2.

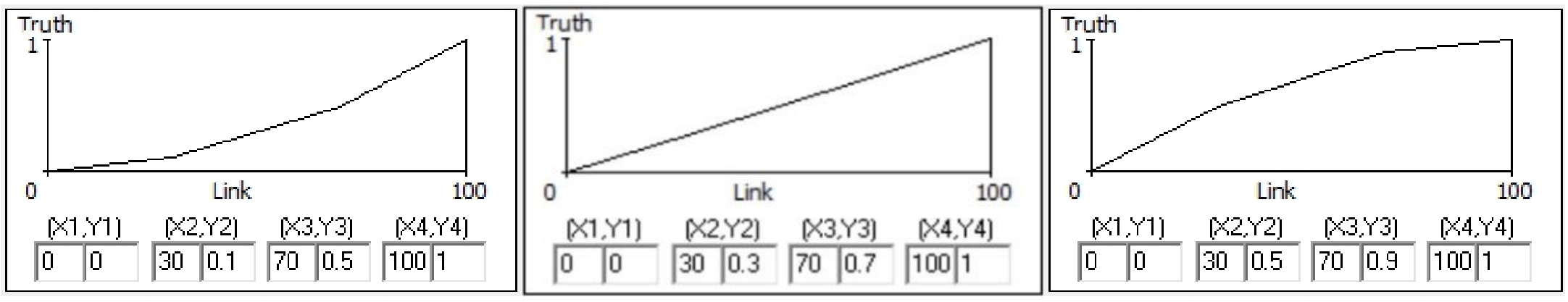

**Fig. 2.** Definition of the (from the left) *Concave*, *Linear* and *Convex* membership functions for the fuzzy variable *Strong_Relationship* in the universe of discourse *Percent of the maximum link value between objects*.

These factors were selected because they correspond to the primary determinants of the forces that drive agent movement during the simulation. NO and LD define the structural complexity of the layout problem. VF governs how strongly objects (agents) respond to truth-value deficits in the linguistic patterns. MD and ML shape how distance and link intensities are interpreted in the fuzzy-logic framework, influencing how attraction and repulsion are computed.

The combination of all factor levels resulted in: 3 × 3 × 3 × 4 × 3 = 324 experimental conditions.

For every condition, two dependent variables were recorded:

- Classic economic objective function value, expressing the total link–distance cost (EGF),
- NCS introduced in the previous section that measure the layout uniformity degree and reflects how evenly object positions occupy a regular grid overlaying the plane.

This design allowed for detailed analyses of main effects, two-way and higher-order interactions, and patterns essential for formulating parameter-selection recommendations.

### 3.1.2. Simulation Procedure

Each of the 324 experimental conditions was evaluated through a structured simulation workflow consisting of randomized initialization, iterative agent movement, and performance tracking. The procedure was as follows:

(1) Generation of relationship matrices.

For every condition, object pairs were assigned link values drawn randomly from the integer range 1–9. Link density was controlled by specifying the proportion of non-zero entries in the matrix. This allowed the experiment to emulate a broad spectrum of realistic layout scenarios, ranging from sparse to highly interconnected systems.

(2) Execution of simulation replications.

To account for the stochastic nature of wandering-agent dynamics, and specifically the random initialization of agent positions, each experimental condition was replicated 30 times. This number of replications ensured robust estimation of performance and mitigated the influence of unfavorable random starts.

(3) Movement simulation.

Each simulation run consisted of 100 algorithmic steps, during which agents (objects) adjusted their positions on the 800 × 800 pixel plane, based on (for details see Grobelny & Michalski, 2025):

- the degree to which linguistic patterns were satisfied,

– the resulting truth-value deficits,

– the virtual forces derived from linguistic patterns and dependent on membership functions, and

– the interacting effects of attraction and repulsion.

These dynamics were parameterized using the independent factors and served to emulate the progressive optimization of the layout.

(4) Performance recording.

After the simulation steps were completed, the resulting object coordinates were used to compute both the classic GF and NCS. For each condition, the best value obtained across the 100 replications was retained for subsequent analysis. These best-case results enabled direct comparison across conditions and informed the construction of the recommendation tables presented later in the study.

## 3.2. Results

The purpose of this section is to examine how the five manipulated factors: NO, LD, VF, MD, and ML, influence the two dependent performance measures: EGF and the uniformity-based NCS measure. Because both measures respond to the complex interplay between structural properties of the layout problem and the linguistic-pattern parameters guiding the virtual-force algorithm, the analyses were conducted in two complementary stages.

First, full five-way factorial ANOVAs were performed to quantify the general sensitivity of both dependent variables to all main effects and higher-order interactions across the entire 324-condition design. These analyses provide a comprehensive overview of the extent to which each factor contributes to economic performance and uniformity, and they reveal the presence of broad, statistically robust interaction structures.

Second, because significant high-order interactions can obscure interpretable patterns and practical tuning strategies, additional restricted three-way ANOVAs (VF × MD × ML) were performed. These analyses isolate the effects of VF, MD, and ML within representative combinations of NO and LD, allowing the identification of clear response patterns, region-specific behavior, and optimal parameter settings. Together, these two stages offer both a global and context-specific understanding of how virtual-force–driven linguistic-pattern algorithms behave under different problem characteristics.

### 3.2.1. Overview of the Full Factorial Five-way Analyses of Variance

The five-way ANOVAs (Table 1) reveal that virtually all manipulated factors exert statistically significant influences on both EGF and NCS, with the majority of two-way and higher-order interactions also reaching significance. This confirms that both dependent variables are highly sensitive to the combined effects of structural complexity (NO and LD) and the characteristics of the linguistic-pattern interpretation (VF, MD, and ML).

Table 1. Five-way full factorial analyses of variance results for EGF and uniformity NCS.

| Factor | Classic EGF | | | Uniformity NCS | |
|---|---|---|---|---|---|
| | **df** | **F** | **p** | **F** | **p** |
| Problem size (NO) | 2 | 167 001 | <0.0001* | 562 | <0.0001* |

| Link density (LD) | 2 | 194 096 | <0.0001* | 1330 | <0.0001* |
|---|---|---|---|---|---|
| Virtual force (VF) | 2 | 89.1 | <0.0001* | 366 | <0.0001* |
| Dist Mem Fun (MD) | 3 | 17.1 | <0.0001* | 27.6 | <0.0001* |
| Lnk Mem Fun (ML) | 2 | 35.4 | <0.0001* | 14.2 | <0.0001* |
| NO×LD | 4 | 39 631 | <0.0001* | 143 | <0.0001* |
| NO×VF | 4 | 13.5 | <0.0001* | 65.4 | <0.0001* |
| NO×MD | 6 | 1.7 | 0.123 | 3.4 | 0.0021* |
| NO×ML | 4 | 5.3 | 0.0003* | 0.49 | 0.740 |
| LD×VF | 4 | 32.9 | <0.0001* | 151 | <0.0001* |
| LD×MD | 6 | 14.3 | <0.0001* | 13.0 | <0.0001* |
| LD×ML | 4 | 39.3 | <0.0001* | 47.5 | <0.0001* |
| VF×MD | 6 | 4.5 | 0.0001* | 88.6 | <0.0001* |
| VF×ML | 4 | 2.8 | 0.0255* | 59.4 | <0.0001* |
| MD×ML | 6 | 6.9 | <0.0001* | 18.3 | <0.0001* |
| NO×LD×VF | 8 | 39.5 | <0.0001* | 54.4 | <0.0001* |
| NO×LD×MD | 12 | 8.0 | <0.0001* | 88.5 | <0.0001* |
| NO×LD×ML | 8 | 9.1 | <0.0001* | 25.6 | <0.0001* |
| NO×VF×MD | 12 | 3.3 | 0.0001* | 8.7 | <0.0001* |
| NO×VF×ML | 8 | 2.2 | 0.0274* | 16.6 | <0.0001* |
| NO×MD×ML | 12 | 3.9 | <0.0001* | 2.2 | 0.0101* |
| LD×VF×MD | 12 | 5.5 | <0.0001* | 10.8 | <0.0001* |
| LD×VF×ML | 8 | 4.7 | <0.0001* | 4.2 | <0.0001* |
| LD×ML×MD | 12 | 7.5 | <0.0001* | 11.8 | <0.0001* |
| VF×MD×ML | 12 | 1.0 | 0.452 | 3.4 | <0.0001* |
| NO×LD×VF×MD | 24 | 2.4 | 0.0001* | 22.3 | <0.0001* |
| NO×LD×VF×ML | 16 | 2.8 | 0.0002* | 4.5 | <0.0001* |
| NO×LD×MD×ML | 24 | 5.8 | <0.0001* | 9.1 | <0.0001* |
| NO×VF×MD×ML | 24 | 2.5 | 0.0001* | 7.1 | <0.0001* |
| LD×VF×MD×ML | 24 | 1.8 | 0.0115* | 5.1 | <0.0001* |
| NO×LD×VF×MD×ML | 48 | 2.8 | <0.0001* | 7.4 | <0.0001* |
| Error | 9396 | | | | |

However, despite the presence of many significant interactions, the main effects remain informative for illustrating the general direction and magnitude of factor influence. The large F-values observed for NO and LD, in particular, indicate that these structural factors dominate the overall variance, while the remaining factors modulate the optimization outcome in more specific ways.

Because the full interaction space in a five-factor design is too complex to visualize concisely, we present main-effect plots to provide readers with an intuitive summary of the overarching trends. These plots do not replace the formal factorial analyses but rather complement them by presenting the dominant response patterns that hold across most conditions. Accordingly, the following figures illustrate how changes in NO, LD, VF, MD, and ML generally affect the EGF and the uniformity-based NCS, offering a clear visual interpretation of the consistent trends underlying the statistical results.

### Classic EGF plots for main effects

The main-effect plots presented in Figs. 3-7 derived from the five-way ANOVA provide a clear overview of how each manipulated factor influences the economic goal function (EGF).

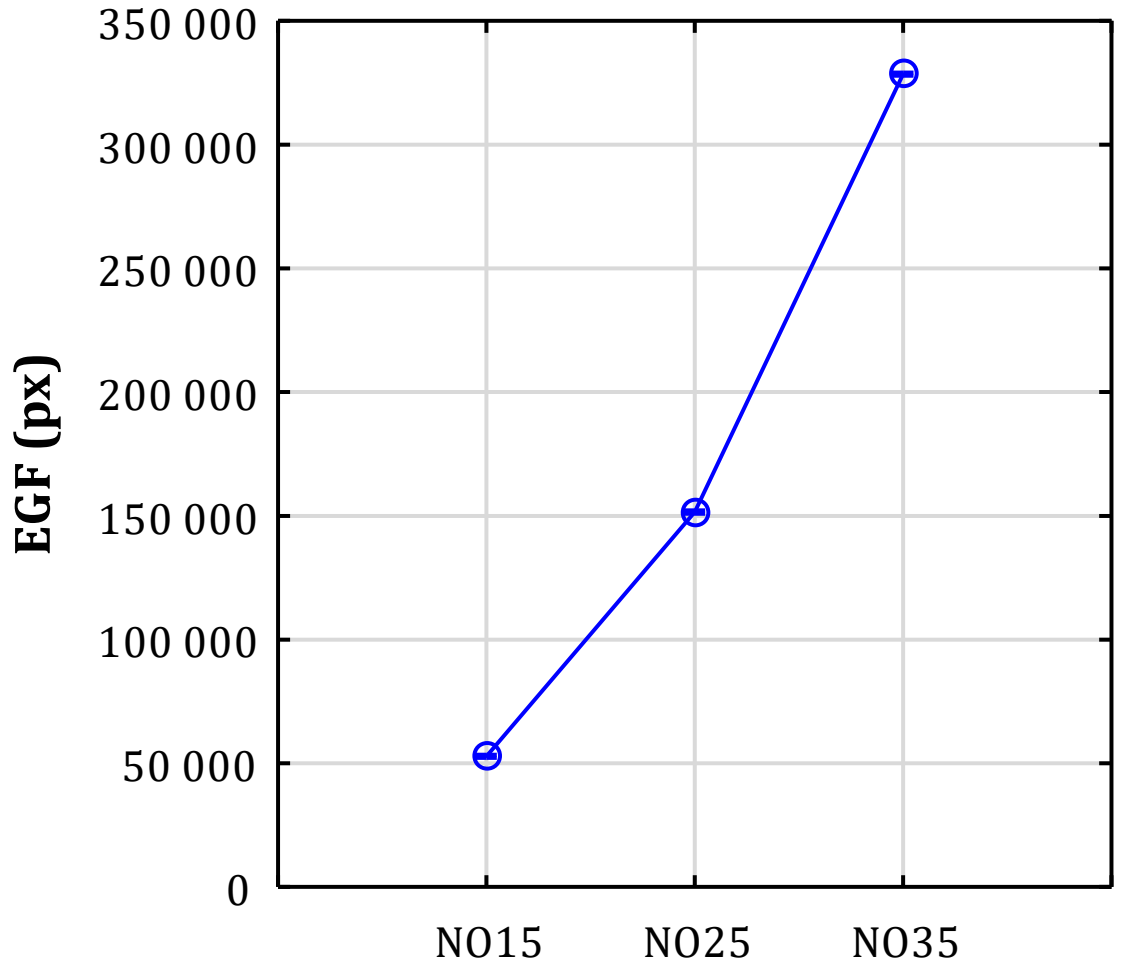

Fig. 3. Main effect of NO on EGF. $F(2, 9396) \approx 1.7 \times 10^5$, $p < 0.0001$. Vertical bars denote 0.95 confidence intervals.

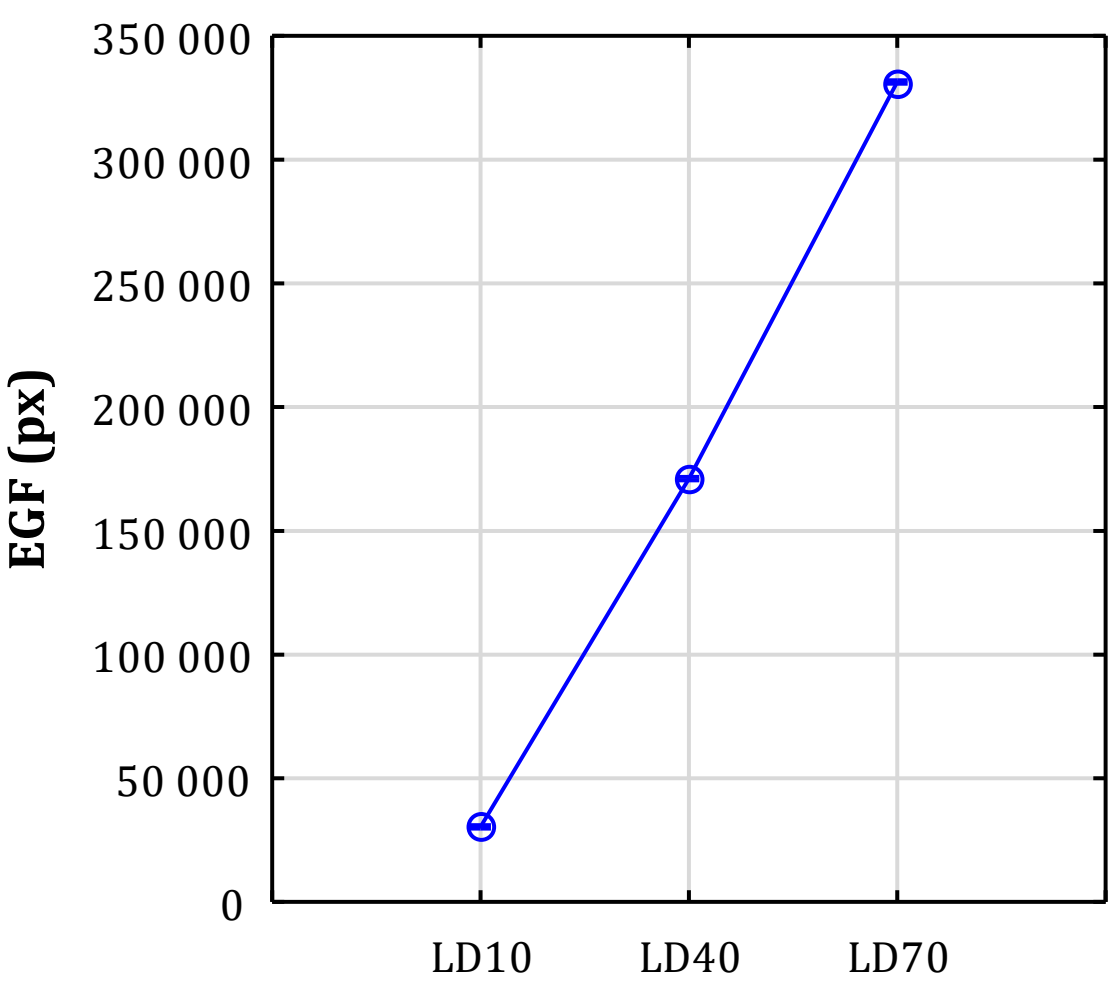

Fig. 4. Main effect of LD on EGF. $F(2, 9396) \approx 1.9 \times 10^5$, $p < 0.0001$. Vertical bars denote 0.95 confidence intervals.

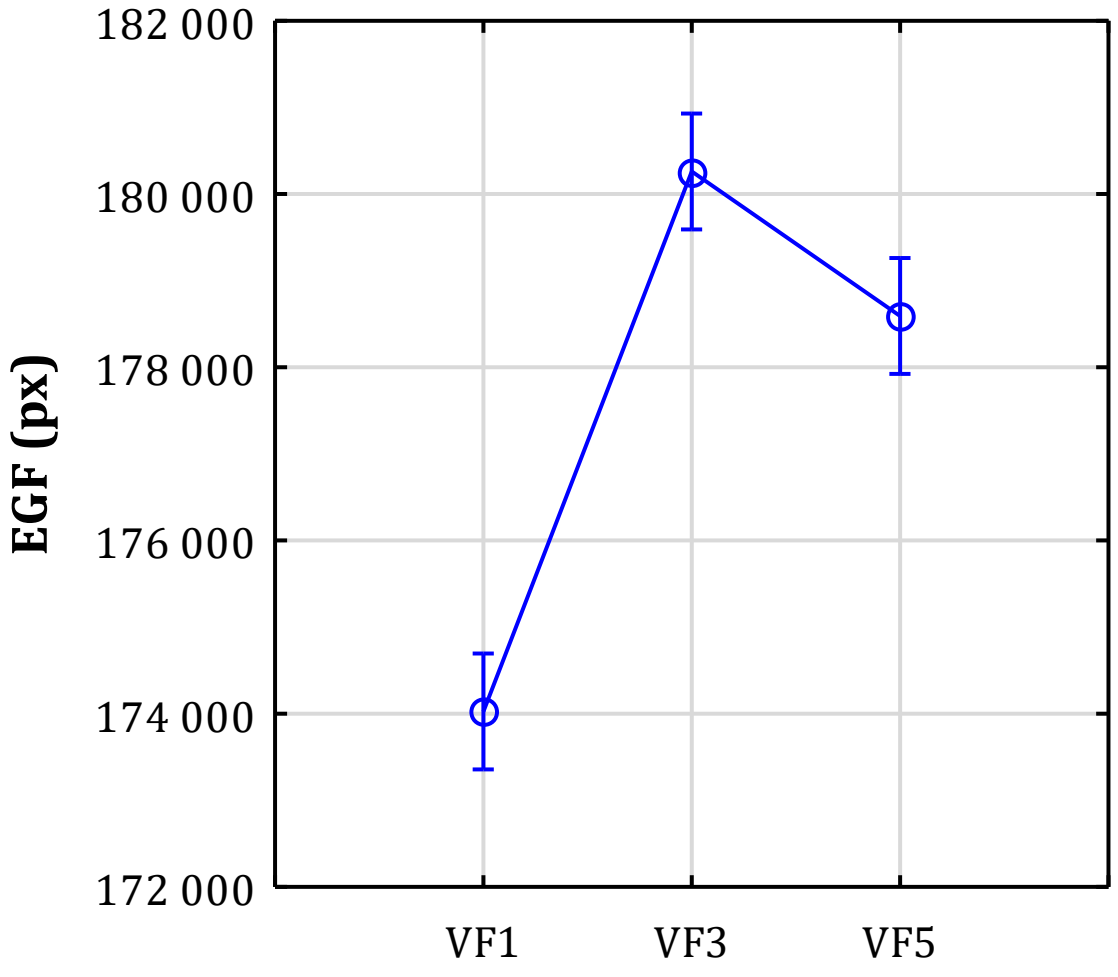

Fig. 5. Main effect of VF on EGF. $F(2, 9396) = 89$, $p < 0.0001$. Vertical bars denote 0.95 confidence intervals.

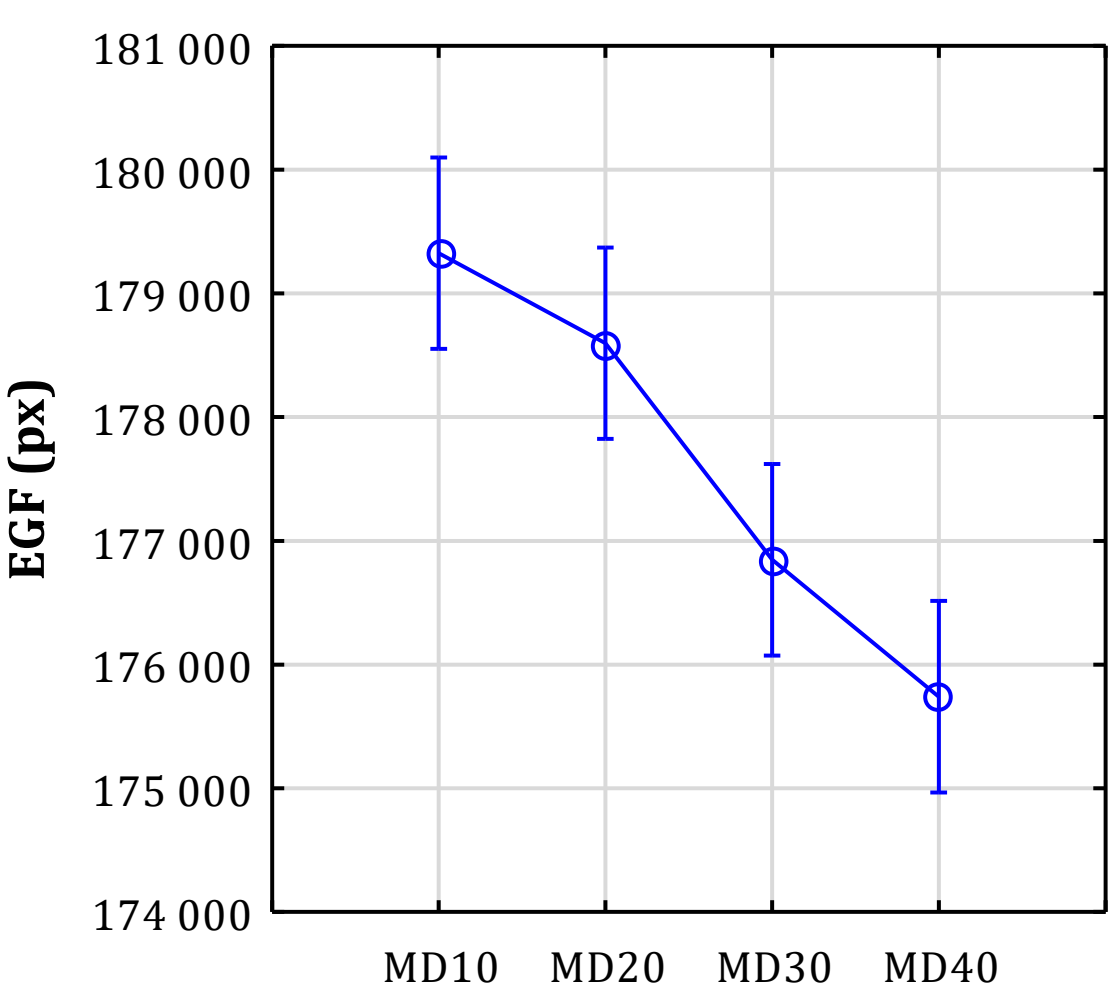

Fig. 6. Main effect of MD on EGF. $F(3, 9396) = 17$, $p < 0.0001$. Vertical bars denote 0.95 confidence intervals.

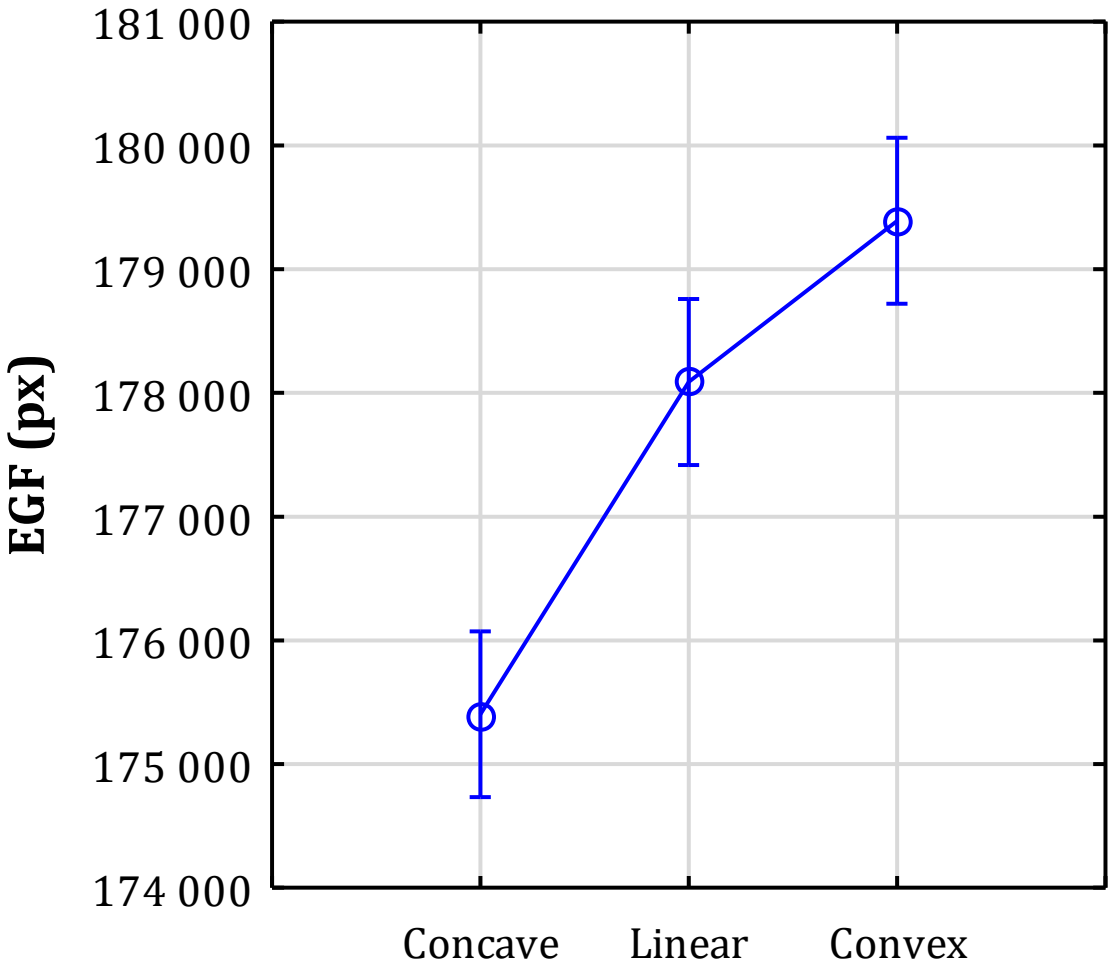

Fig. 7. Main effect of ML on EGF. $F(2, 9396) = 35$, $p < 0.0001$. Vertical bars denote 0.95 confidence intervals.

Among all factors, the NO exhibits one of the most dominant trends. As NO increases from 15 to 35 items, the EGF rises steeply, reflecting the rapid growth in cumulative link–distance cost as problem size expands (Fig 3). This pattern is accompanied by an extremely large F-value ($F(2, 9396) \approx 1.7 \times 10^5$, $p < 0.0001$), confirming NO as a central structural determinant of EGF.

An even stronger effect is observed for LD. Increasing LD from sparse (10%) to dense (70%) produces the largest increase in EGF across any factor, as shown by the substantial range on the y-axis of the LD (Fig 4). This aligns with expectations: denser connectivity multiplies the number of weighted distances contributing to the cost function. Correspondingly, LD produces the largest F-value in the entire ANOVA ($F(2, 9396) \approx 1.9 \times 10^5$, $p < 0.0001$), identifying it as the single strongest main effect.

In contrast to these structural factors, the VF coefficient shows only a modest variation in its plot (Fig 5). Although VF is statistically significant ($F(2, 9396) = 89$, $p < 0.0001$), the differences between its levels (1%, 3%, 5%) span only a narrow range, suggesting that VF acts primarily as a fine-tuning parameter rather than a dominant driver of economic performance. The algorithm responds systematically to changes in VF, but its influence is small relative to the effects of NO and LD.

The MD exerts an smaller effect than VF, with visible mean separations among the four MD levels. Its associated ANOVA statistic ($F(3, 9396) = 17$, $p < 0.0001$) confirms that the steepness of the distance-membership decline reliably influences the resulting layouts (Fig 6).

The ML has the medium (between VF and MD) main effect among the algorithmic/linguistic parameters. *Concave*, *Linear*, and *Convex* transformations of link values all yield significantly different EGF outcomes, as confirmed by the high F-value ($F(2, 9396) = 35$, $p < 0.0001$) and clear separation of mean values in Fig 7. This demonstrates that the manner in which link strengths are represented in the fuzzy-logic framework meaningfully alters the system's sensitivity to EGF gradients and thereby influences layout quality.

Taken together, these main-effect results indicate that NO and LD overwhelmingly determine the baseline magnitude of the economic cost, essentially defining the difficulty level of each scenario. In contrast, VF, MD, and ML act as algorithmic shaping mechanisms, allowing researchers and practitioners to adjust how the virtual-force system interprets distances and link intensities. This strong and consistent sensitivity of the economic objective to changes in both structural and algorithmic parameters, provides a solid foundation for the more detailed interaction analyses presented in the subsequent series of 3-way ANOVA analyses.

### NCS plots for main effects

The NCS main-effect results from Figs. 8-12 indicate that all five factors exert statistically reliable influences on spatial uniformity, with NO emerging as the dominant driver of between-factor differences. Across the three NO levels, the Fig. 8 shows clear and sizable separation among means, confirming that NCS is highly sensitive to the number of objects. It is consistent with the notion that altering the count of placed items systematically changes how evenly the plane is occupied. It occurs despite the normalization denominator in NCS. This is probably due to a number of significant interactions of NO with other examined factors.

LD also produces a robust and clearly patterned change in NCS across its levels (Fig. 9). As the connectivity pattern becomes sparser or denser, the figure shows visible shifts in the average proportion of occupied vs. available grid cells with the maximum NCS value for the LD level of 70%. This is consistent with the idea that the global structure of the interaction network conditions how evenly objects can be distributed in space. Relative to NO, the LD band is slightly wider and and still substantial.

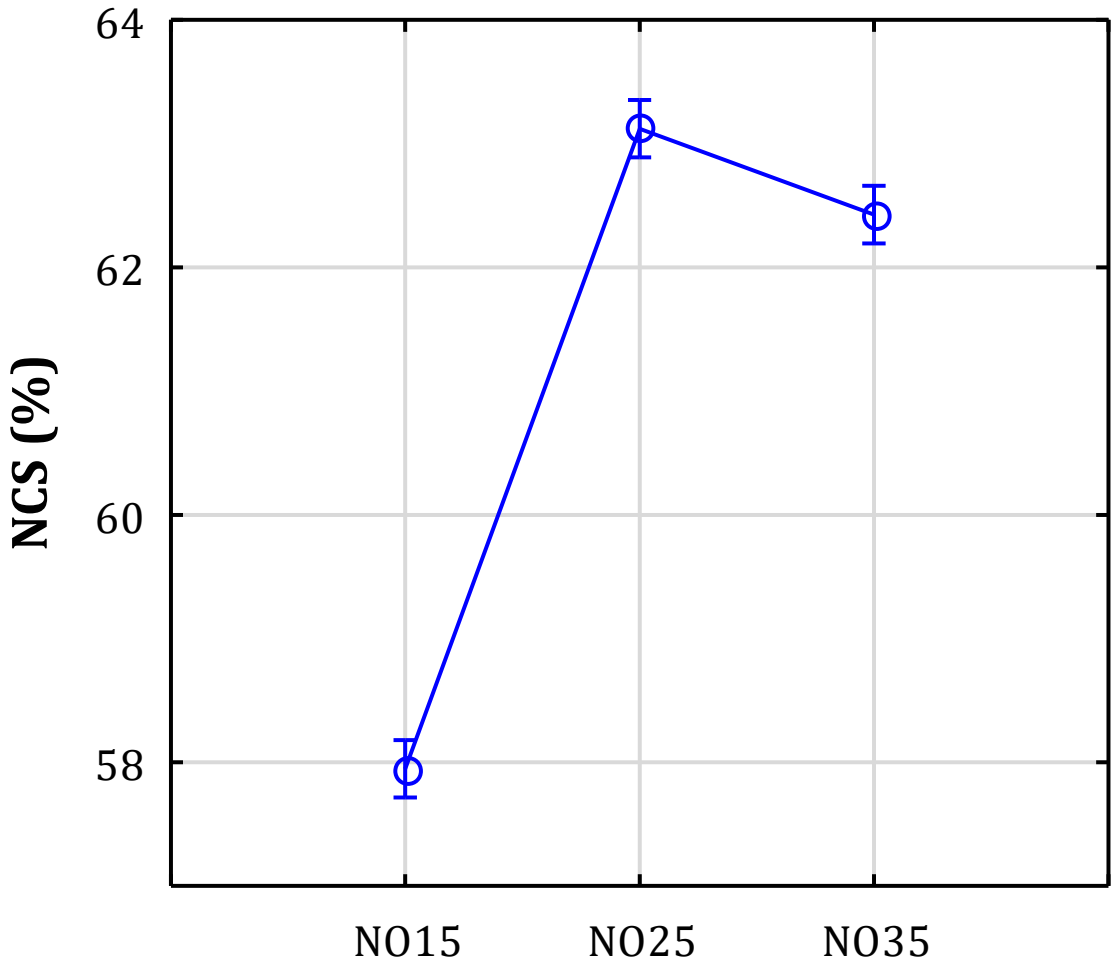

Fig. 8. Main effect of NO on NCS. $F(2, 9396) = 562$, $p < 0.0001$. Vertical bars denote 0.95 confidence intervals.

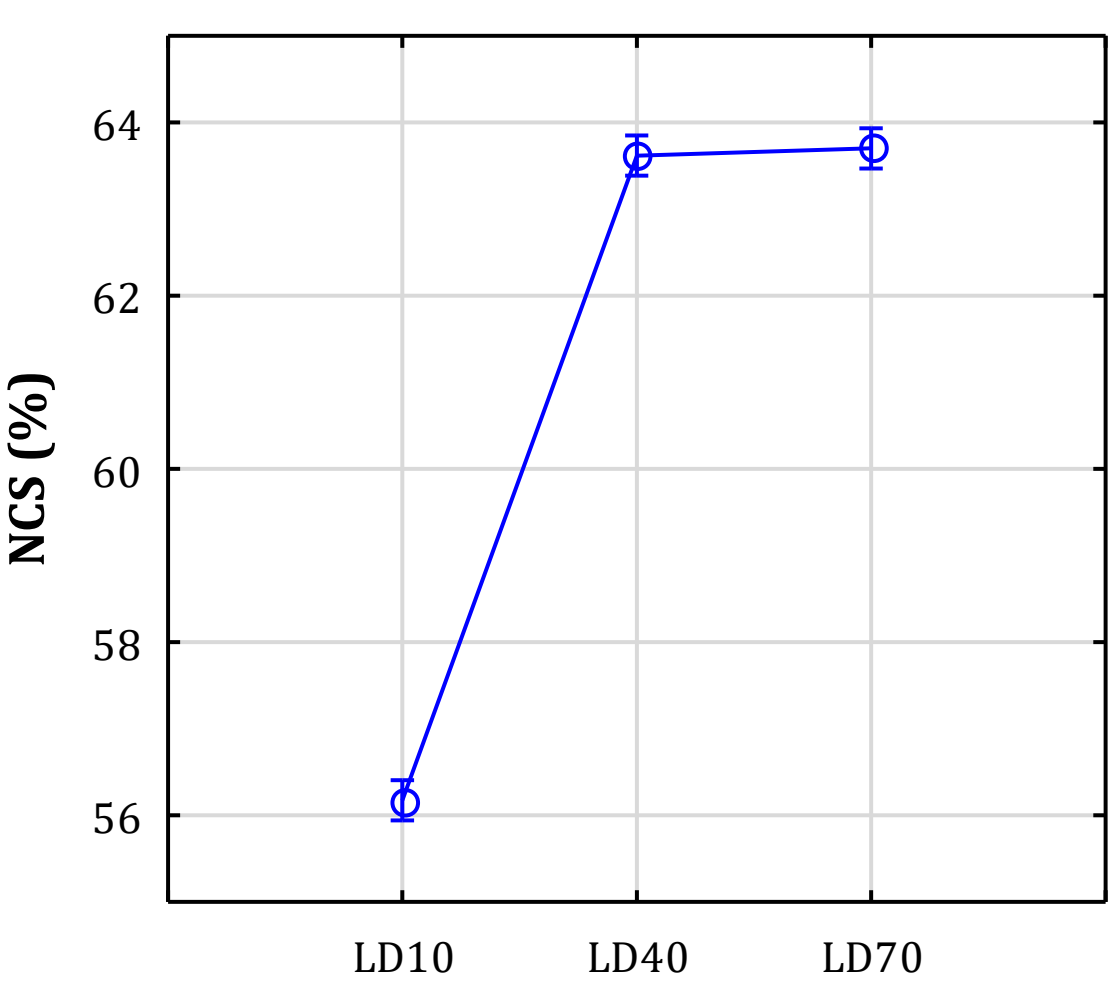

Fig. 9. Main effect of LD on NCS. $F(2, 9396) = 1330$, $p < 0.0001$. Vertical bars denote 0.95 confidence intervals.

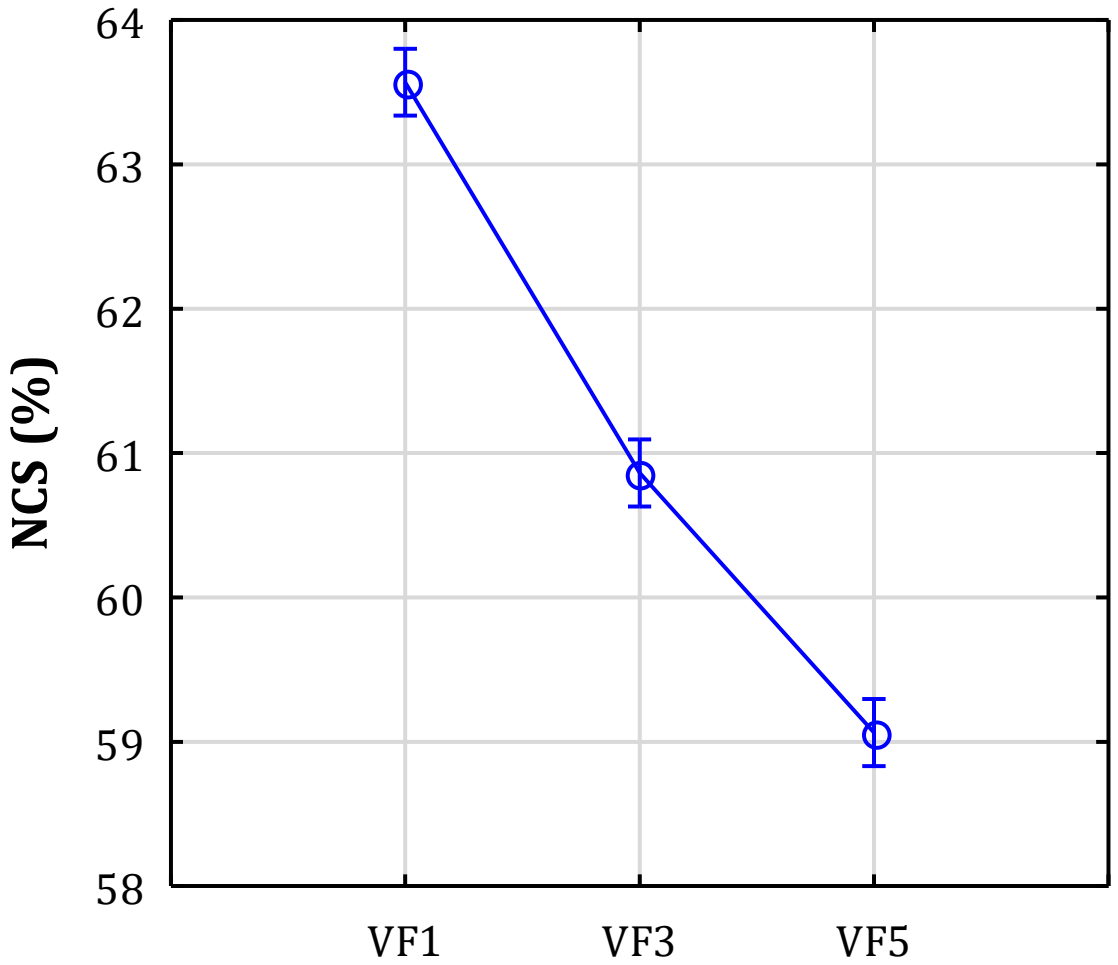

Fig. 10. Main effect of VF on NCS. $F(2, 9396) = 366$, $p < 0.0001$. Vertical bars denote 0.95 confidence intervals.

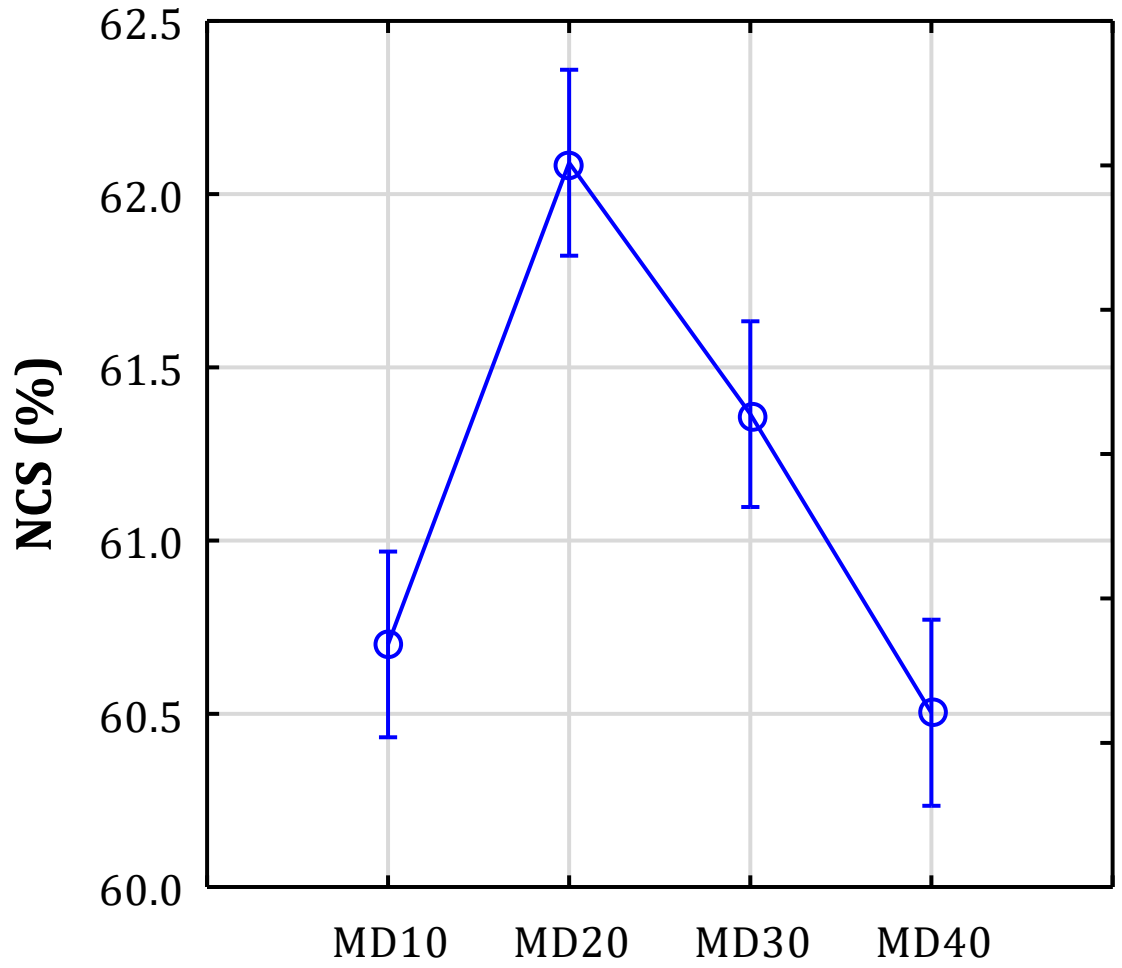

Fig. 11. Main effect of MD on NCS. $F(3, 9396) = 28$, $p < 0.0001$. Vertical bars denote 0.95 confidence intervals.

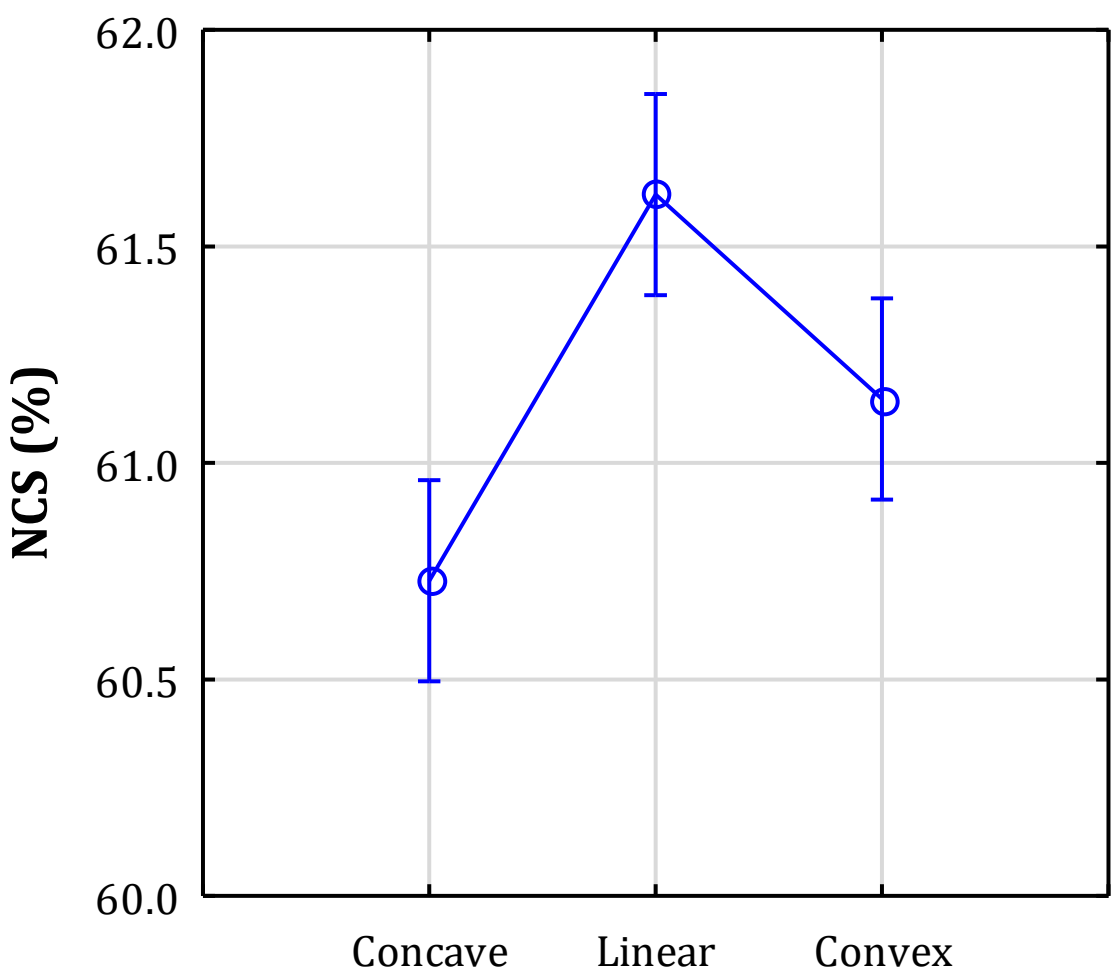

Fig. 12. Main effect of ML on NCS. $F(2, 9396) = 14$, $p < 0.0001$. Vertical bars denote 0.95 confidence intervals.

For the algorithmic parameters, VF shows a smaller, but still systematic, spread across its three levels (Fig. 10), indicating that the overall force scale fine-tunes how the system approaches uniform occupancy without overturning the broader influence of NO and LD. Similarly, both MD (Fig. 11) and ML (Fig. 12) display clear separation among their profiles (four and three levels, respectively), confirming that the linguistic-pattern mapping of distances and link values meaningfully modulates uniformity beyond the structural constraints imposed by NO and LD. In each case, the NCS figure presents consistent differences across the levels, supporting reliable main-effect conclusions.

Taken together, the main-effect panels establish that structural factors (NO, LD) set the baseline difficulty for both uniformity (NCS) and economic objective (EGF), while algorithmic/linguistic factors (VF, MD, ML) shape performance within those structural constraints.

#### 3.2.2. A series of ANOVAs for combinations of NO and LD

While the global ANOVAs from previous section establish that all factors and many of their interactions significantly influence performance, such complexity makes direct interpretation challenging for practitioners seeking to optimize algorithm performance. To provide a more detailed and practically oriented understanding, a second analytical stage was conducted using nine restricted three-way ANOVAs (VF × MD × ML). Each of these was performed within a specific combination of NO (15, 25, or 35) and LD (10%, 40%, or 70%).

This targeted approach allows the interaction patterns among VF, MD, and ML to be examined under controlled problem conditions, revealing how these algorithmic mechanisms behave in small, medium, and large problems, and under sparse, moderate, or dense levels of structural complexity and network connectivity. Because these combinations closely resemble typical scenarios encountered in real facility layout problems, the resulting interaction patterns directly inform parameter-selection recommendations.

Supplementary material A (Figs. A1–A18) presents the results of all 18 three-way ANOVAs (NO × LD for both EGF and NCS) in the form of interaction plots for both dependent variables across all nine

NO–LD combinations. These visualizations make it possible to identify the most effective membership-function shapes, distance sensitivities, and virtual-force coefficients for each problem category. The patterns extracted from these interactions form the empirical basis for the recommendations presented in Table 2. Cells marked with × in the Min **E** row denote factor-level configurations that achieved the minimum average value of the economic objective EGF, whereas cells marked with × in the Max **U** indicate configurations that yielded the maximum average uniformity measured by the proposed NCS.

Since NO and LD are fixed structural attributes of the given optimization problem, Table 2 is intended to be used as a lookup tool: given a problem characterized by (NO, LD), one simply selects the recommended levels of the three tunable parameters, the experimental factors, i.e., VF, MD, and ML.

Table 2. Recommendation table for choosing the algorithm parameters based on the two series (for EGF and NCS) of nine three-way (VF × MD× ML) ANOVAs arising from 3(NO) × 3 (LD).

| **Prob. Size (NO)** | **Link Dens (LD) %** | **Optimized Goal Function Type** | **Virtual force (VF)** | | | **Distance Mem Fun (MD)** | | | | **Link Mem Fun (ML)** | | |
|---|---|---|---|---|---|---|---|---|---|---|---|---|
| | | | **1** | **3** | **5** | **10** | **20** | **30** | **40** | **Linear** | **Concave** | **Convex** |
| **15** | **10** | Min **E** | | | × | × | | | | | | × |
| | | Max **U** | × | | | | | | × | × | | |
| | **40** | Min **E** | | × | | | | × | | × | | |
| | | Max **U** | × | | | × | | | | | | × |
| | **70** | Min **E** | × | | | × | | | | | × | |
| | | Max **U** | × | | | × | | | | × | | |
| **25** | **10** | Min **E** | | | × | | × | | | | | × |
| | | Max **U** | × | | | × | | | | | | × |
| | **40** | Min **E** | | | × | | | | × | | | × |
| | | Max **U** | × | | | × | | | | | × | |
| | **70** | Min **E** | | | × | | | | × | | | × |
| | | Max **U** | × | | | | | | × | | | × |
| **35** | **10** | Min **E** | × | | | × | | | | | | × |
| | | Max **U** | × | | | | | × | | × | | |
| | **40** | Min **E** | | | × | × | | | | × | | |
| | | Max **U** | × | | | | | | × | | | × |
| | **70** | Min **E** | | | × | × | | | | | | × |
| | | Max **U** | × | | | | | | × | × | | |

Cells marked with × in the Min **E** refer to factor-level configurations that yielded the lowest average values of the economic objective. Cells marked with × in Max **U** rows indicate configurations that delivered the highest average uniformity, measured by NCS.

# 4. Discussion

While the preceding analyses quantify how individual factors and their interactions influence economic performance and spatial uniformity, the statistical significance alone does not immediately translate into actionable guidance for practitioners. The recommendation matrix derived from the restricted three-way ANOVAs provides such guidance, but its effective use requires understanding the mechanisms through which VF, MD, and ML jointly shape layout outcomes. To bridge this gap, the next subsection offers an interpretation of the recommendation patterns, explaining why specific parameter configurations favor economic compaction, spatial dispersion, or a balanced compromise. This interpretation connects the empirical structure of the results with practical decision-making in real-world layout design.

## 4.1. Interpretations of Recommendations

The "×" markers for Min E, and Max U from the recommendation matrix (Table 2) reveal a highly structured and consistent pattern across all eighteen ANOVAs. To facilitate a clear understanding of how these empirical findings translate into practical guidance, we interpret the recommendation matrix from three perspectives: minimization solely of EGF (Min E), maximization of uniformity measured by NCS (Max U), and simultaneous optimization of EGF and NCS.

### 4.1.1. Minimization of EGF

Across nearly all problem configurations, the strongest predictor of low economic objective value is a high virtual force. Rows corresponding to Min E are dominated by settings with VF = 5%, with occasional reinforcement from VF = 3%. This confirms that intensifying the virtual attraction between linked objects systematically reduces the sum of link-weighted distances, generating compact, economically efficient layouts. The effect of the MD is almost monotonic. For Min E rows, the recommended MD levels cluster overwhelmingly at MD10 and MD20, the two steepest membership function. Lower MD levels promote spatial contraction, reinforcing the influence of VF and supporting economical configurations. The ML factor plays a refinement role. The Min E rows most often mark *Convex* ML (and, in several configurations, *Linear* ML), which strengthen the significance of connections relative to distances, thereby amplifying the push toward compact, low-economic objective layouts.

Taken together, the Min E rows suggest a straightforward combination, which is stable across all problem sizes and link densities: VF = 5%, MD = 10 or 20, ML = *Convex* (with *Linear* acceptable in some problem cases).

### 4.1.2. Maximization of Uniformity (NCS)

The Max U rows of Table 2 consistently favor low VF values, with VF = 1% emerging as the dominant recommendation. Weaker virtual forces allow the objects to be relatively less close to each other in final layouts. The recommendations for the MD function shift decisively toward the upper end of the scale (the least steep decreasing functions). MD = 30 and MD = 40 repeatedly appear in Max U rows. These MD levels promote dispersion and help distribute objects across a greater number of grid cells, increasing uniformity. For the ML, the Max U recommendations almost universally favor *Linear* ML, with occasional support from *Concave* ML. These shapes do not favor the significance of links in relation to distances, and thus help maintain a more homogeneous spatial structure.

Also in the case of the uniformity-oriented perspective the recommendation is relatively clear: VF = 1%, MD = 30 or 40, ML = *Linear* (with *Concave* as an admissible variant).

### 4.1.3. Simultaneous Optimization of EGF and NCS

Since NO and LD parameters describe the problem itself, the table is interpreted as a decision guide under fixed constraints. When the intention is to optimize both criteria simultaneously, the pair of rows (Min E and Max U) must be consulted for the given (NO, LD) and identify overlaps between recommended VF, MD, ML levels, or adjacent levels that minimize conflict between the objectives.

Since the VF and MD factors reveal a clear antagonistic relationship, i.e., Min E prefers VF = 5% and MD = 10–20, while Max U prefers VF = 1% and MD = 30–40. Thus, VF = 3% and MD = 20–30 serve as structurally consistent compromise choices, appearing near the *transition* zones between the two objectives. Although VF = 3% rarely dominates either objective, it is centrally located between the contractions induced by VF = 5% and the expansions induced by VF = 1%, making it a rational compromise setting visible in several intermediate cases.

The ML factor exhibits less conflict, that is, Max U overwhelmingly prefers *Linear*, Min E frequently chooses *Convex*, but sometimes *Linear*. Therefore, for multi-objective use, ML = *Linear* is generally the safest non-conflicting choice, because it remains acceptable for Min E in several (NO, LD) configurations and is the dominant recommendation for Max U.

As a result of these considerations, the general prescription for simultaneous Min E & Max U optimization can be formulated as VF = 3%, MD = 20 or 30, and ML = *Linear*. This compromise balances contraction and dispersion, yielding layouts that do not strongly favor one criterion at the expense of the other. Table 2 shows that these intermediate settings often appear when a problem does not strongly lean toward either extreme outcome.

## 4.2. Comparison with Other Algorithms

To situate LP-Alinks within the broader methodological landscape, we directly compared it with competing methods. Using the parameter settings from Table 2, we conducted a series of experiments for each of the analyzed problem types in order to examine the behavior of LP-Alinks approach in comparison with the fundamental classical procedures used in this domain. In each experiment, we searched for optimal solutions using the methods of eigenvector-based Drezner (Drezner, 1980, 1987a), MDS (Torgerson, 1952, 1958) and NmMDS (Kruskal, 1964a, 1964b; Shepard, 1962a, 1962b), and we computed for these solutions the two previously defined evaluation criteria: the economic and uniformity objective.

### 4.2.1. Simulations Results

For our algorithm, solutions were obtained by running 30 independent simulations for each examined problem, using the parameter settings specified in Table 2. In every experiment, we selected the solution with the best value of the objective function. Since the Drezner and MDS methods are analytical and yield a single deterministic solution, the values of both criteria depend solely on the definitions of the corresponding measures. For NmMDS, we used a version allowing random initialization, and selected the best solution from 100 repetitions. The resulting objective function values obtained in this comparative study are summarized in Table 3.

Table 3. Comparison of economic and uniformity (NCS) objectives with solutions provided by Drezner's, Multidimensional Scaling (MDS), and NonMetric MDS (NmMDS) algorithms.

| **Prob. Size (NO)** | **Link Dens (LD) %** | **Drezner** | | **MDS** | | **NmMDS** | | **LP-Alinks** | |
|---|---|---|---|---|---|---|---|---|---|
| | | **Min E** | **Max U %** | **Min E** | **Max U %** | **Min E** | **Max U %** | **Min E** | **Max U %** |
| **15** | **10** | 2961 | 44 | 8227 | 44 | 7635 | 63 | 5210 (×)<br>7315 | 50<br>69 (×) |
| | **40** | 34099 | 44 | 51362 | 63 | 46826 | 69 | 43654 (×)<br>44187 | 75<br>75 (×) |
| | **70** | 75117 | 50 | 111369 | 63 | 99321 | 69 | 91370 (×)<br>96911 | 69<br>69 (×) |
| **25** | **10** | 5119 | 28 | 24633 | 60 | 19956 | 64 | 11136 (×)<br>14565 | 52<br>68 (×) |
| | **40** | 103330 | 56 | 142348 | 68 | 150556 | 68 | 129003 (×)<br>129842 | 68<br>76 (×) |
| | **70** | 116037 | 20 | 283040 | 60 | 268609 | 72 | 242955 (×)<br>256239 | 72<br>80 (×) |
| **35** | **10** | 32696 | 30 | 52252 | 50 | 56240 | 64 | 38991 (×)<br>38388 | 67<br>69 (×) |
| | **40** | 155768 | 33 | 291570 | 64 | 304391 | 64 | 249568 (×)<br>269086 | 56<br>67 (×) |
| | **70** | 368058 | 44 | 619347 | 61 | 620697 | 67 | 481701 (×)<br>561750 | 47<br>69 (×) |

The comprehensive evaluation these test scenarios reveals clear structural differences among the examined methods. Across all configurations, Drezner's technique demonstrates a pronounced dominance in minimizing the economic objective function. It consistently yields the lowest values, reflecting its strong bias toward reducing total costs. However, this economic efficiency is systematically accompanied by very low uniformity. The method generates highly irregular layouts, and the deterioration in spatial regularity becomes more severe as the number of objects and link density increase. Although Drezner remains the benchmark for pure cost minimization, its geometric instability limits its usefulness in applications requiring clear or balanced spatial designs.

Both MDS and NmMDS produce layouts with substantially higher uniformity, which results from their focus on preserving relative distances or dissimilarities rather than minimizing connection costs. Consequently, they yield significantly higher economic objective values. NmMDS consistently outperforms classical MDS due to its non-metric formulation, which enables more flexible embeddings and better captures the underlying dissimilarity structure. Nonetheless, as problem size and link density increase, the uniformity of both techniques gradually declines.

Compared with Drezner, NmMDS achieves considerably better uniformity and a more balanced trade-off between geometric clarity and economic performance. However, its economic objective values remain markedly higher, confirming that NmMDS does not primarily optimize cost.

The most significant findings concern the LP-Alinks method. It consistently attains the highest uniformity across all scenarios and outperforms both MDS and NMDS in terms of economic objective values. Notably, these values remain close to the best results produced by Drezner, even when tested under conditions specifically designed to maximize uniformity.

These results indicate that LP-Alinks successfully integrates geometric regularity and economic efficiency within a single optimization framework. Unlike Drezner, it prevents excessive distance collapse. Unlike MDS-type methods, it directly incorporates cost-related structure into the optimization. The method demonstrates strong robustness and scalability, maintaining high uniformity and competitive economic objective values even as problem complexity increases.

Importantly, LP-Alinks preserves spatial regularity under challenging settings with large numbers of objects and high link density, where the uniformity of MDS and NmMDS degrades substantially. Its strong performance in difficult parameter regimes suggests that LP-Alinks achieves a deeper and more intrinsic balance between economic and geometric criteria.

### 4.2.2. Qualitative Analysis of Extreme Cases

Let us consider and visually analyze the most extreme case of non-uniformity among the results in Table 3. In Fig. 13(a) we show a scatter plot for a task with with 25 objects (5×5 matrix) and 70% random link density, for which the worst NCS of 20% was obtained by the Drezner's approach (EGF = 116 037). This figure illustrates an important feature of Drezner's method resulting directly from its optimization assumptions: weakly connected objects are strongly repelled from the rest, causing excessive sparsity in certain regions and local overcrowding in others. Despite the best economic objective value, the practical usefulness of such a layout is questionable, even when we disregard the physical dimensions of the objects. Fig. 13(b), in turn, presents the same problem after performing LP-Alinks algorithm (EGF = 256 239; NCS = 80%). By contrast, the layout obtained with our method is much clearer and could be physically realized with only minor position adjustments to account for object sizes.

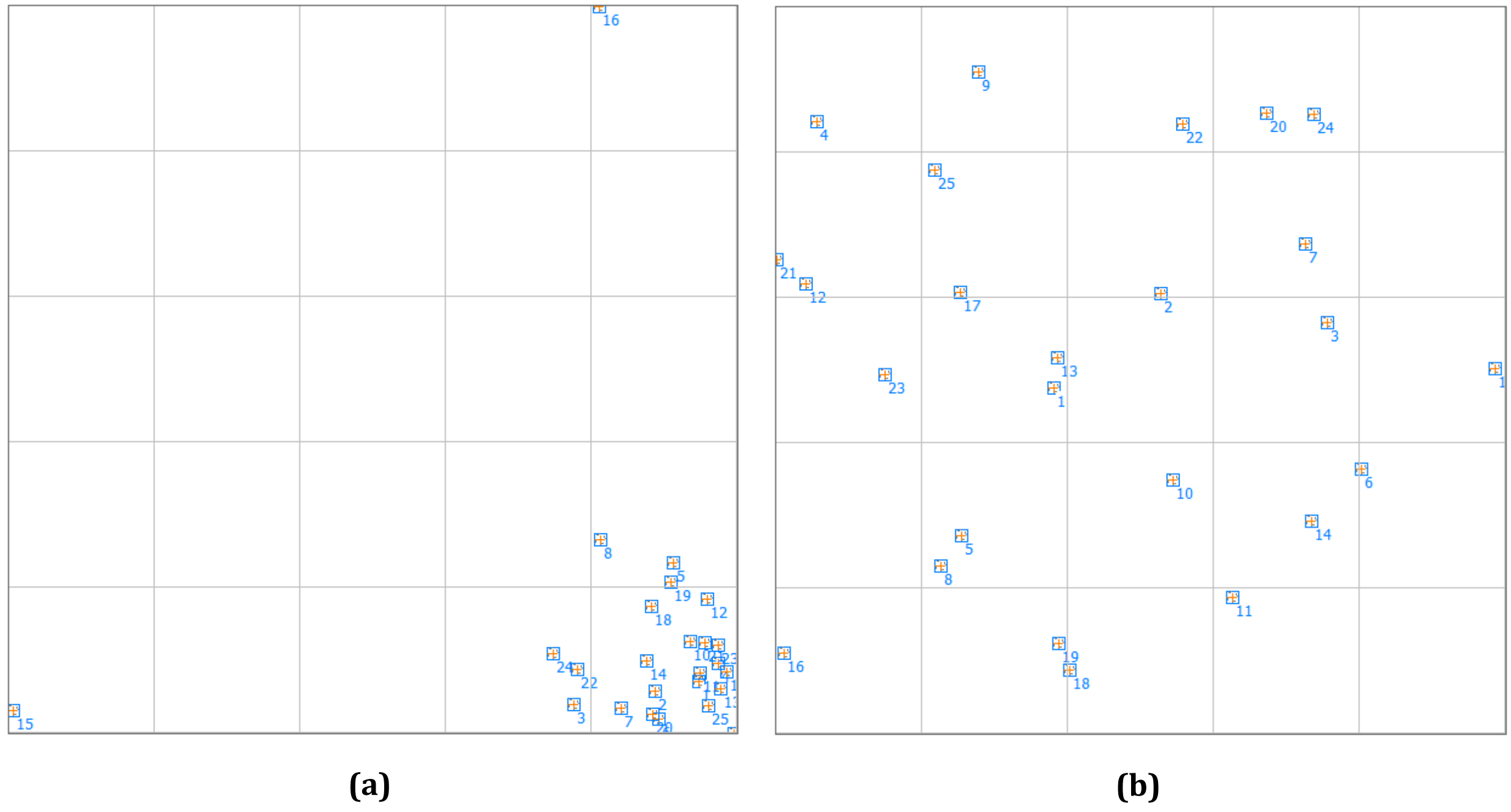

**Fig. 13.** Solutions to the problem including 25-elements with 70% link density. (a) Drezner's layout with EGF = 116 037; NCS = 20%; (b) LP-Alinks with EGF = 256 239; NCS = 80%.

Another interesting example is presented in Fig. 14(a) and shows the best NCS result obtained using Drezner's method for the problem involving 25-elements with 40% link density (EGF = 103 330; NCS = 56%). The best solution for the same problem but obtained by LP-Alinks method while using parameters for maximizing uniformity is demonstrated in Fig. 14(b) (EGF = 129 842; NCS = 76%).

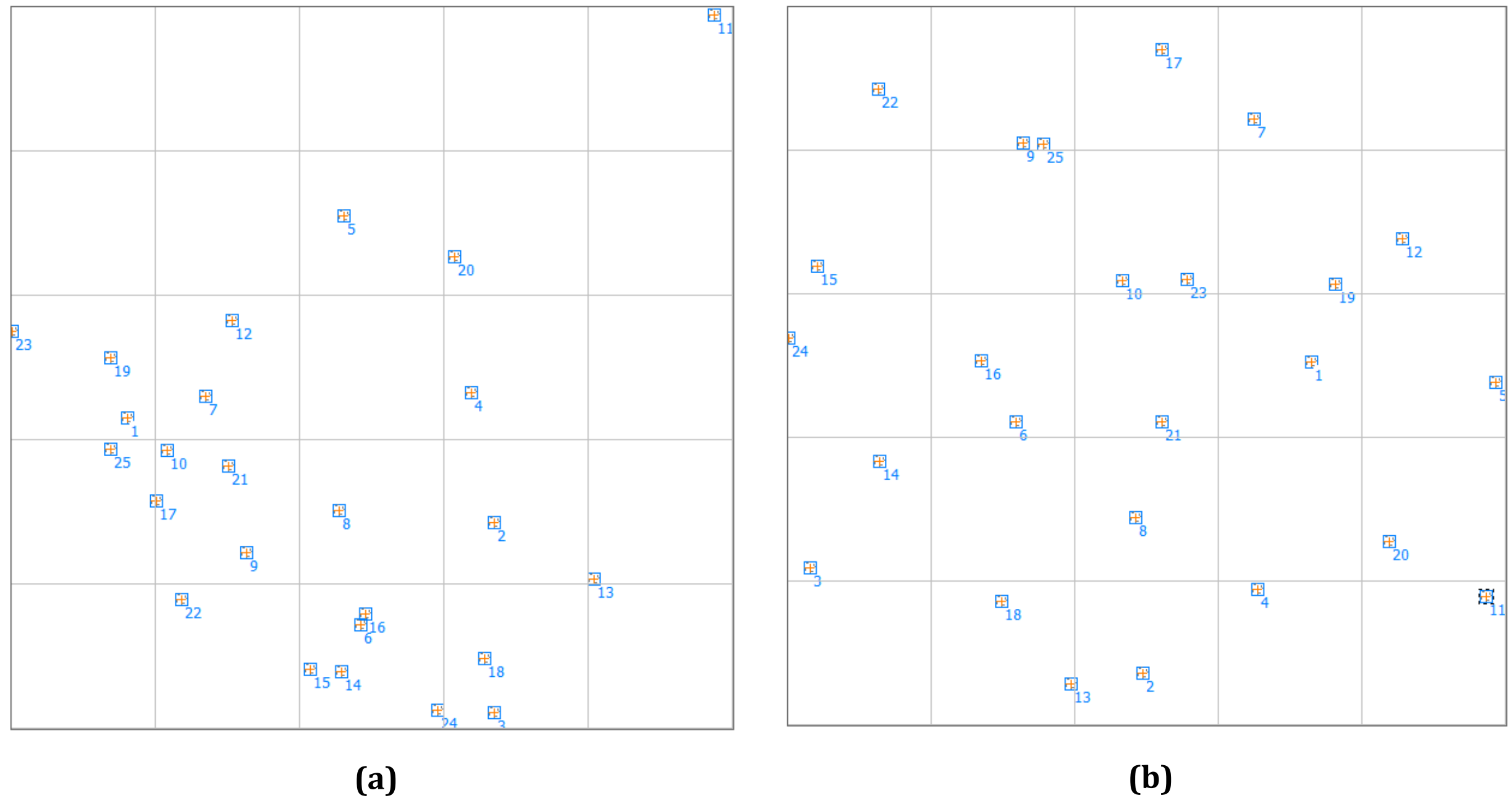

**Fig. 14.** Solutions to the problem including 25-elements with 40% link density. (a) Drezner's layout with EGF = 103 330; NCS = 56%; (b) LP-Alinks with EGF = 129 842; NCS = 76%.

In this example Drezner-based layout is reasonably readable, though more crowded than the solution proposed by LP-Alinks. The reduced uniformity in Fig. 14(a) results mainly from the substantial displacement of a single object (number 11). This displacement also explains the difference in EGF values, although the difference is not drastic here. This result suggests that our approach provides layout structures of comparable quality with respect to economic function. While confirming this observation requires more systematic analysis, the experimental results presented here demonstrate the capability of the LP-Alinks algorithm to generate layouts with practically desirable properties.

The compromise achieved between economic and uniformity criteria values ensures that the resulting layouts can be applied almost directly to the design of real-world systems under practical project conditions. Importantly, the generation of layout solution by our algorithm follows a linguistic-pattern framework, in which both relational constraints and evaluation criteria can incorporate the knowledge, experience, and intuition of a designer-expert.

## 4.3. Theoretical Contributions and Practical Implications

### 4.3.1. Theoretical Contributions

This work advances the theory of layout optimization by articulating and empirically investigating a two-criterion framework that clarifies the fundamental trade-off between the economic objective and spatial uniformity. By explicitly framing these objectives and demonstrating that LP-Alinks can approximate a balanced region of the trade-off surface, the paper provides a conceptual model for reasoning about cost-geometry interactions and a methodological template for multi-objective exploration.

The introduction of the NCS formalizes spatial evenness as a mathematically explicit objective alongside the classical flow-distance cost. Theoretically, the NCS may operate as a geometric regularizer that discourages the collapsed, highly clustered layouts often resulting from cost-only optimization, thereby improving the legibility and interpretability of scatter-plot designs. Because the NCS is computed directly on the 2D grid used by designers, it remains simple, interpretable, and compatible with continuous scatter-plot formats. Furthermore, by scaling the evaluation grid with problem size, the score stays roughly invariant to the number of objects, which avoids misleading fluctuations due only to $N$ and enables fair comparisons across diverse instances in factorial studies.

Five-way ANOVAs show that NO and LD are responsible for the majority of the variance in both the EGF and NCS. These structural factors define the feasible envelope of performance, while algorithmic and linguistic factors (VF, MD, and ML) shape the specific outcomes within that envelope. This hierarchy explains why main-effect plots remain informative even when complex interactions are present. Across representative conditions, the interactions between these parameters form stable patterns; specifically, the NCS responds more directly to membership definitions and force scaling than the economic objective does. A practical equilibrium point is identified at VF = 3% with MD in the mid-range and *Linear* ML, representing a functional balance between attraction-induced compaction and dispersion-induced coverage. Consistent separations among MD and ML levels indicate that these functions act as geometry controllers, where steeper membership functions promote compaction to benefit the EGF, while *Linear* ML support the dispersion necessary for a high NCS.

Moreover, we unify classic and LP methods within one landscape: Drezner directly minimizes path length (lowest EGF, but poor NCS), MDS and NmMDS minimize stress to preserve dissimilarities (higher NCS, higher EGF) while LP-Alinks embeds domain semantics by LP yielding balanced cost–uniformity outcomes. This synthesis clarifies each method's role in the design space and explains the observed trade-offs.

The study's evidence strengthens a theoretical position that facility layouts should be studied as multi-criteria geometric embeddings under constraints. By offering a simple, scale-aware uniformity metric and a semantics-driven dynamical optimizer, the framework provides building blocks for unifying FLP, spatial statistics, and visual analytics. The theoretical payoff is a portable recipe for other spatial design problems (e.g., control-panel layout, HCI dashboards, or multi-facility siting) where economic rationality and geometric clarity must be co-optimized.

### 4.3.2. Practical Implications

Practical implications are mostly related to the recommendation table and its analysis provided in the discussion section. In practice, given a specific optimization problem, the user identifies NO and LD which define a unique pair of rows in Table 2 (Min E and Max U rows). Then, levels of VF, MD, ML that optimize the required objective can be read off. This makes the table a direct prescriptive tool and no search over parameter space is required, and no additional model calibration is needed.

In a brief summary, the table reveals three coherent recommendations:

- Economic optimality measured by EGF (Min E): VF = 5%, MD = 10–20, ML = *Convex*.
- Uniformity optimality measured by NCS (Max U): VF = 1%, MD = 30–40, ML = *Linear*.
- Dual-objective compromise (Min E and Max U): VF = 3%, MD = 20–30, ML = *Linear*.

These patterns are consistent across all eighteen experimental ANOVAs and provide a transparent, practitioner-ready guide for setting algorithmic parameters under any of the three optimization intents.

### 4.4. Limitations and Future Research

While LP-Alinks demonstrates a favorable balance between EGF and NCS, several limitations of the present study create opportunities for future work. First, the scope of instances is constrained. In our simulation experiments we used random matrices with predefined sizes and link densities, as specified in Table 2. Although this facilitates reproducibility, it does not fully capture structures common in real design problems (e.g., community modularity, hierarchical organization, or spatial embedding). Consequently, external validity beyond the tested distributions still needs to be established. A systematic expansion toward structured synthetic data and real-world datasets (e.g., facility cells, production lines, communication topologies) is required to demonstrate generalization.

Furthermore, we assessed solutions using the economic objective EGF and a grid-based uniformity measure (NCS). While transparent, these metrics capture only part of what practitioners regard as readability and feasibility. Future work could incorporate other measures, such as minimum-spacing violations, aisle widths, or maintenance access, and explicitly treat layout quality as a multi-facet construct. Additionally, parameter sensitivity and stochasticity merit deeper analysis. LP-Alinks performance depends on parameter settings and both LP-Alinks and NmMDS are affected by random initialization. Our policy of selecting the best out of 100 runs per task mitigates variance but may overstate typical outcomes.

Finally, the current experiments deliberately omit physical and regulatory constraints (e.g., object sizes, minimum clearances, routing corridors, safety/ergonomic standards), which simplifies comparison but limits direct deployability. Because LP-Alinks already encodes expert knowledge through linguistic patterns, it is a natural vehicle for integrating such constraints so that solutions are feasible-by-construction and require fewer post-edits. In parallel, human-in-the-loop studies are needed to measure usability, inter-expert variability, and the learning curve associated with crafting and refining linguistic rules. Together, these steps would strengthen real-world adoption.

## 5. Conclusions

This study shows that LP-Alinks is a practical early-stage layout method that co-optimizes economic cost (EGF) and spatial uniformity (NCS) within one, linguistically driven framework. By expressing design rules as linguistic patterns and converting their truth deficits into virtual forces, the approach embeds expert intent directly in the optimization dynamics and produces readable scatter-plot layouts while remaining competitive on cost.

Across the full factorial design, NO and LD dominate variability and therefore set the feasible envelope of attainable performance. Within that envelope, VF, MD, and ML reliably tune outcomes. In detailed analyses, VF × MD × ML interactions exhibit stable patterns; a balanced configuration: VF = 3%, MD in the mid-range, and ML = *Linear* often reconciles compaction (cost) with coverage (uniformity), while stronger curvature settings favor cost and milder ones favor uniformity.

The method comparison places classical techniques and the proposed approach on a common footing. Drezner remains preferred for pure cost minimization but typically yields low uniformity; MDS/NmMDS improve uniformity at the price of higher cost. LP-Alinks offers the most balanced trade-off, achieving consistently high NCS with EGF markedly better than MDS/NmMDS and often close to Drezner, and it maintains this behavior as complexity and density increase.

Finally, the recommendation table (Table 2) converts these findings into direct, prescriptive guidance by mapping (NO, LD) to effective VF, MD, ML settings for economic, uniformity, or balanced goals. In sum, LP-Alinks advances layout optimization by integrating domain semantics with multi-objective search, delivering interpretable, scalable layouts and ready-to-apply parameter choices for scenarios where economic rationality and geometric clarity must be satisfied together.

## Conflict of Interest

The authors have no competing interests to declare that are relevant to the content of this article.